\documentclass[aps,preprintnumbers,twocolumn,amsmath,amssymb,prl,showpacs]{revtex4}
\usepackage{bbm}
\usepackage{mathrsfs}
\usepackage{amsmath}
\usepackage{graphicx}
\usepackage{epstopdf}

\usepackage{color}

\newcommand{\be}{\begin{equation}}
\newcommand{\ee}{\end{equation}}
\newcommand{\bea}{\begin{eqnarray}}
\newcommand{\eea}{\end{eqnarray}}
\newcommand{\bes}{\begin{split}}
\newcommand{\ees}{\end{split}}

\renewcommand{\vec}[1]{\mathbf{#1}}

\newcommand{\tr}{\operatorname{Tr}}

\newcommand{\bs}{\boldsymbol}

\begin{document}
\title{A Memory of Majorana Modes through Quantum Quench}
\author{Ming-Chiang Chung$^{1,2}$, Yi-Hao Jhu$^3$, Pochung Chen$^{2,3}$,
  Chung-Yu Mou$^{2,3,4}$ and Xin Wan$^5$ }
\affiliation{$^1$ Physics Department, National Chung-Hsing University, Taichung, 40227, Taiwan}
\affiliation{$^2$Physics Division, National Center for Theoretical Science, Hsinchu, 30013, Taiwan}
\affiliation{$^3$Physics Department, National Tsing Hua University, Hsinchu, 30013, Taiwan}
\affiliation{$^4$Institute of Physics, Academia Sinica, Taipei 11529, Taiwan}
\affiliation{$^5$Zhejiang Institute of Modern Physics, Zhejiang University, 
Hangzhou, 310027, P. R. China}

\begin{abstract}
We study the sudden quench of a one-dimensional p-wave superconductor
through its topological signature in the entanglement spectrum. The
long-time evolution of the system and its topological characterization
depend on a pseudomagnetic field ${\bs R}_{\mbox{\text eff}}(k)$,
which connects both the initial and the final Hamiltonians, hence
exhibiting a memory effect. In particular, we explore the robustness of
the Majorana zero-mode associated with the entanglement cut in the
topologically nontrivial phase and identify the parameter space in
which the mode can survive in the infinite-time limit. 
\end{abstract} 

\date{\today}
\maketitle 

{\it Introduction--}The implementation of a quantum computer requires
fault-tolerant quantum information processing. Conventional quantum
error correction codes~\cite{Shor} demand an error threshold that is
still beyond the reach of the current technology. Alternatively, one
may exploit exotic topological excitations that obey non-Abelian
braiding statistics to encode quantum information, which would then be
robust against local perturbations. Named after Ettore
Majorana~\cite{Majorana}, Majorana zero-modes seem to be the easiest
to construct among the family of objects that realize non-Abelian
statistics. A prototypical model for Majorana zero-modes is the
one-dimenaional (1D) p-wave supercoductor~\cite{Kitaev}, in which different
topological phases can be characterized by a topological
$\mathcal{Z}_2$ index. Topological non-trivial phases can be
identified by the presence of zero-energy Majorana edge modes at open
boundaries, which may be realized at the interface of superconductors
with either topological insulators~\cite{FuKane} or semiconductors
with strong spin-orbit coupling~\cite{SauDasSarma,vonOppen}. Recent
experimental progress~\cite{MFExp} further fuels the interest in the
preparation and manipulation of Majorana zero-modes.

To control the Majorana zero-modes for braiding or computing one needs
to dynamically change the experimental parameters, such as the gate
voltage in a wire network~\cite{Alicea} or the magnetic flux in a
hybrid Majorana-transmon device~\cite{Hyart}. This motivated the study
of the out-of-equilibrium dynamics of systems with Majorana modes at
the ends~\cite{slowQuench}, which found that topology could induce
anomolous defect production that could cause quantum decoherence, when
a system was adiabatically driven through a quantum critical point.  On
the other hand, a sudden quench (not necessarily on a topological
system) gives rise to the question whether the system can undergo
relaxation to an equilibrium state upon a change of
parameters. Interestingly, recent studies also showed that the quench
dynamics of integrable systems has a memory effect: it reaches a
steady state depending strongly on the initial condition
\cite{Rigol,MingMiguelAnibal}. However, most studies along that line
only concern the bulk properties with few exceptions~\cite{Patel}.  In
order to use Majorana zero-modes as robust quantum information
carriers, it is, hence, of great interest to study the stability of
the Majorana edge modes in quench dynamics.

An alternative avenue that connects Majorana zero-modes and quantum
information is quantum entanglement. The common entanglement
measurement is the von Neumann entropy of a subsystem $A$: $S_A = -\tr
\rho_A \log_2 \rho_A$, where $\rho_A = {\tr}_B |\Psi_{A \cup B}
\rangle \langle \Psi_{A \cup B} |$ is the reduced density matrix after
tracing out the environment $B$ from the whole system $A \cup B$. For
a topological system the size-independent constant of the entanglement
entropy is related to the total quantum dimension
\cite{KitaevPreskill}, which can be used to detect topological
order. Nevertheless, more information is revealed in the entanglement
spectrum, i.e. the eigenvalues of the entanglement Hamiltonian ${\cal
  H}_A^{ent}$, whose thermodynamic entropy at ``temperature'' $T = 1$
is equivalent to the entanglement entropy~\cite{Haldane}. Under
certain circumstances one can view, at least on the low-energy scale,
the entanglement Hamiltonian on the subsystem $A$ with the open
boundaries as the deformed real-space Hamiltonian that perserves the
topological information, hence the presence of zero-energy Majorana
edge modes can be detected by a corresponding degeneracy in the
entanglement spectrum; more precisely, a pair of doubly degenerate
eigenvalues of $1/2$ in the one-particle entanglement
spectrum~\cite{RyuHatsugai02,RyuHatsugai06, MJCY11, MJCM13}. This
provides a realiable measurement of the Majorana edge modes.

In this Letter we fuse the two subjects by exploring the quench
dynamics of Majorana zero-modes of a 1D p-wave superconductor under
the entanglement measurement process, thereby offer a quantum
information perspective of the manipulation of topological systems and
the robustness of the Majorana zero-modes under sudden quench. We find
that the topology of the infinite-time behavior can be determined by
the properties of a pseudomagnetic field $\vec{R}_{\mathbf{eff}}$,
which connects both the initial and the final Hamiltonians, hence
exhibiting a memory effect. In general, a quench across any phase
boundary will not give rise to Majorana zero-modes. Surprisingly, 
the quench within the same topologically nontrivial phase may
also lead to the loss of the Majorana modes. We provide a dynamical
phase diagram identifying the parameter space where the initial
Majorana zero-modes can survive in the long-time limit.

{\it 1D p-wave superconductor--}  The 1D p-wave superconducting system
of spinless fermions proposed by Kitaev \cite{Kitaev} is described by the
Hamiltonian  
\be \label{H:pwave} 
   \begin{split}
    H = & \sum_{i}  -t \left (c_i^{+} c_{i+1} +
      c_{i+1}^{+} c_i \right) \\ 
    + & \Delta \left(c_i c_{i+1} + c_{i+1}^{+} c_{i}^{+} \right)  - \mu
    \left( c_{i}^{+} c_i -1/2\right),  
    \end{split} 
\ee 
  with the nearest-neighbor hopping amplitude $t$, superconducting
  gap function $\Delta$, and on-site chemical potential $\mu$.   
  The translational invariant Hamiltonian (\ref{H:pwave}) can be 
 written as
\be  \label{H:R} 
  H = -\sum_{k \in BZ} \left (c_k^{+}, c_{-k} \right ) 
\left[ {\mathbf R}(k) \cdot
  {\boldsymbol \sigma} \right] \left (c_k, c_{-k}^{+} \right )^T , 
\ee 
where ${\boldsymbol \sigma} = (\sigma_x, \sigma_y, \sigma_z)$ are Pauli
matrices, and ${\boldsymbol R}(k) =  (0, -\Delta \sin{k}, t\cos{k} + \mu/2)$ is the pseudomagnetic field. The one-particle energy spectrum is simply 
$\epsilon (k) = \pm 2 R(k) = \pm \sqrt{(2t \cos{k} + \mu)^2 + 4\Delta^2
      \sin^2{k}} $. 
The spinless p-wave superconductor (\ref{H:pwave}) breaks the
time-reversal symmetry but perserves the particle-hole symmetry,
therefore it belongs to the class D according to the classification 
of topological insulators and superconductors; it can be 
characterized by a $Z_2$ topological invariant.

\begin{figure}
\center
\includegraphics[width=7.5cm]{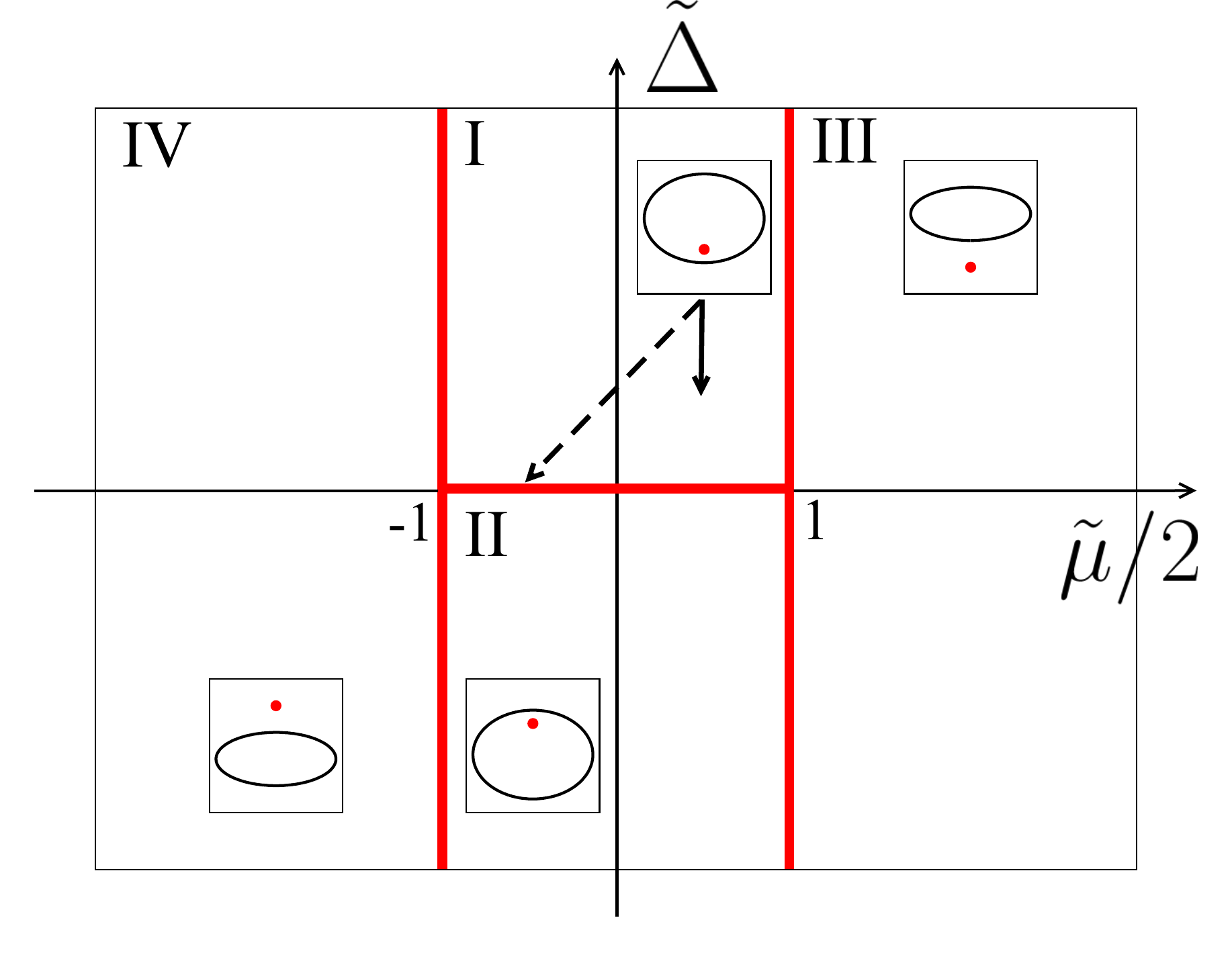}
\caption{(color online) Topological phase diagram of the $p$-wave
superconductor. The coordinates are defined as 
$\tilde{\mu} = \mu/t$ and $\tilde{\Delta} = \Delta/t$. 
Solid arrow: quench process from (0.5,2) to
(0.5,1) [to be discussed in Figs.~\ref{fig3}(a) and \ref{fig4} (b)].
Dashed arrow: quench process from (0.5,2) to (-0.5,0.1) [to be discussed
in Figs.~\ref{fig3}(b) and \ref{fig4}(c)].
Insets: Representative traces of ${\bs R}(k)$ in the $R_y$-$R_z$ plane. 
In the topological phases I and II, ${\bs R}(k)$ encircles the origin 
(red dots), while in the trivial phases III and IV, ${\bs R}(k)$ 
does not encircle the origin. } 
\label{fig1}
\end{figure}

The topological characterization has a simple graphical
interpretation~\cite{RyuHatsugai02}. If the closed loop $\ell$ of
${\boldsymbol R}(k)$ in the $R_y$-$R_z$ plane encircles the origin,
zero-energy edge states exist and the system is in the nontrivial
phase; otherwise, the loop can be continuously deformed to a point
with the bulk gap perserved, hence the system is trivial.  
We plot the phase diagram of the $p$-wave superconductor in
Fig.~\ref{fig1} using half of the dimensionless chemical potential: 
$\tilde{\mu}/2 = \mu/2t$ and
the pairing constant: $\tilde{\Delta} = \Delta/t$ as coordinates. 
For $|\tilde{\mu}/2| < 1$, there are two different
topological nontrivial phases I and II, corresponding to
counterclockwise and clockwise windings of ${\boldsymbol R}(k)$ around
the origin. These two phases cannot be continuously deformed to one
another without closing the bulk gap, so they belong to different
phases. Nevertheless, Majorana zero-modes exist at open ends in both 
phases. On the other hand, the states with $|\tilde{\mu}/2 | > 1$ 
are topologically trivial and no Majorana zero-modes exist 
in phases III and IV.

{\it Entanglement spectrum and Majorana zero-modes-- } 
The reduced density matrix $\rho_A$ can be
calculated by the block correlation function matrix (CFM)
\cite{Review,MingPeschel,GFM} : $\rho_A = \bigotimes_m
\left[\begin{matrix} \lambda_m & 0\\ 0 & 1-\lambda_m \end{matrix}
  \right]$, where $\lambda_m$ are the eigenvalues of the correlation
function matrix $C_{i,j} = \tr \rho {\hat{\bs c}}_i \hat{{\bs
    c}}_j^{+} $ with $\hat{{\bs c}}_i \equiv (c_i, c_i^{+})^T $ and
$i,j$ being sites of the subsystem $A$.  $\lambda_m$ is know as
the one-particle entanglement spectum (OPES).  In the Fourier space 
the correlation function matrix is a $2\times 2$
matrix~\cite{RyuHatsugai06} for the Hamiltonian (\ref{H:R})
\be \label{CFM} C(k) = \frac{1}{2} \left[
  1- \frac{{\bs R}(k) \cdot {\bs \sigma} }{R(k)} \right], 
\ee 
where $k \in (-\pi, \pi] = S^1$. The form of the CFM is the same as
the Hamiltonian (\ref{H:R}) up to a positive normalization and an
additive constant. Therefore, in the topological phases I
and II, the signature of Majorana zero-modes is the degenerate 
eigenvalues $\lambda_m=1/2$ in the OPES. 

The Majorana zero-modes play an important role in the entanglement
between the subsystem $A$ with its environment $B$. We can calculate
the entanglement entropy $E_S$ for the partition as $E_S = \sum_m S_m$
where $S_m = -\lambda_m \log_2 \lambda_m - (1- \lambda_m) \log_2 (1 -
\lambda_m)$.  The pair of Majorana modes with $\lambda_m=1/2$
contribute the maximal entanglement $S_m=1$. Hence they are known 
as the topological maximally-entangled states
(tMES)~\cite{MJCY11,MJCM13}.

{\it Sudden quench across phase boundaries-- } The quench dynamics of
integrable models has become an important topic since such a bulk
system will not be thermalized but reach a steady state described by
generalized Gibbs ensemble (GGE). On the other hand, the Majorana edge
modes are tMES, robust against perturbations, it is interesting to
question how a sudden quench affect the Majorana zero-modes. Naively,
if we quench from a topological phase to a trivial one, the Majorana
modes may evolve into the bulk, mix with bulk modes, and disappear
eventually. What happens, then, if we quench from a trivial phase to a
topological one, or from a topological phase to a different one? Will
the static information in the final Hamiltonian dictate the dynamical
state in the long-time limit?

To answer these questions we perform a sudden quench to the p-wave
superconductor at $t=0$ by switching $\Delta$ and $\mu$ at $t<0$ to
$\Delta'$ and $\mu'$ at $t>0$.  This changes the Hamiltonian from $H$
to $H'$ and, correspondingly, ${\bs R}$ to ${\bs R}'$. We then
calculate the time-dependent OPES $\lambda_m(t)$ by diagonalizing the
time-dependent CFM $G_{ij}(t) = \tr \rho e^{i H' t} \hat{\bs c}_i
\hat{\bs c}_j^{+} e^{- i H' t}$ for $i,j \in A$. We first consider the
quench processes across phase boundaries and focus on 
$\lambda_m$ close to $1/2$ as we are primarily interested in the fate
of the Majorana zero-modes.

\begin{figure}
\center
\includegraphics[width=7.5cm]{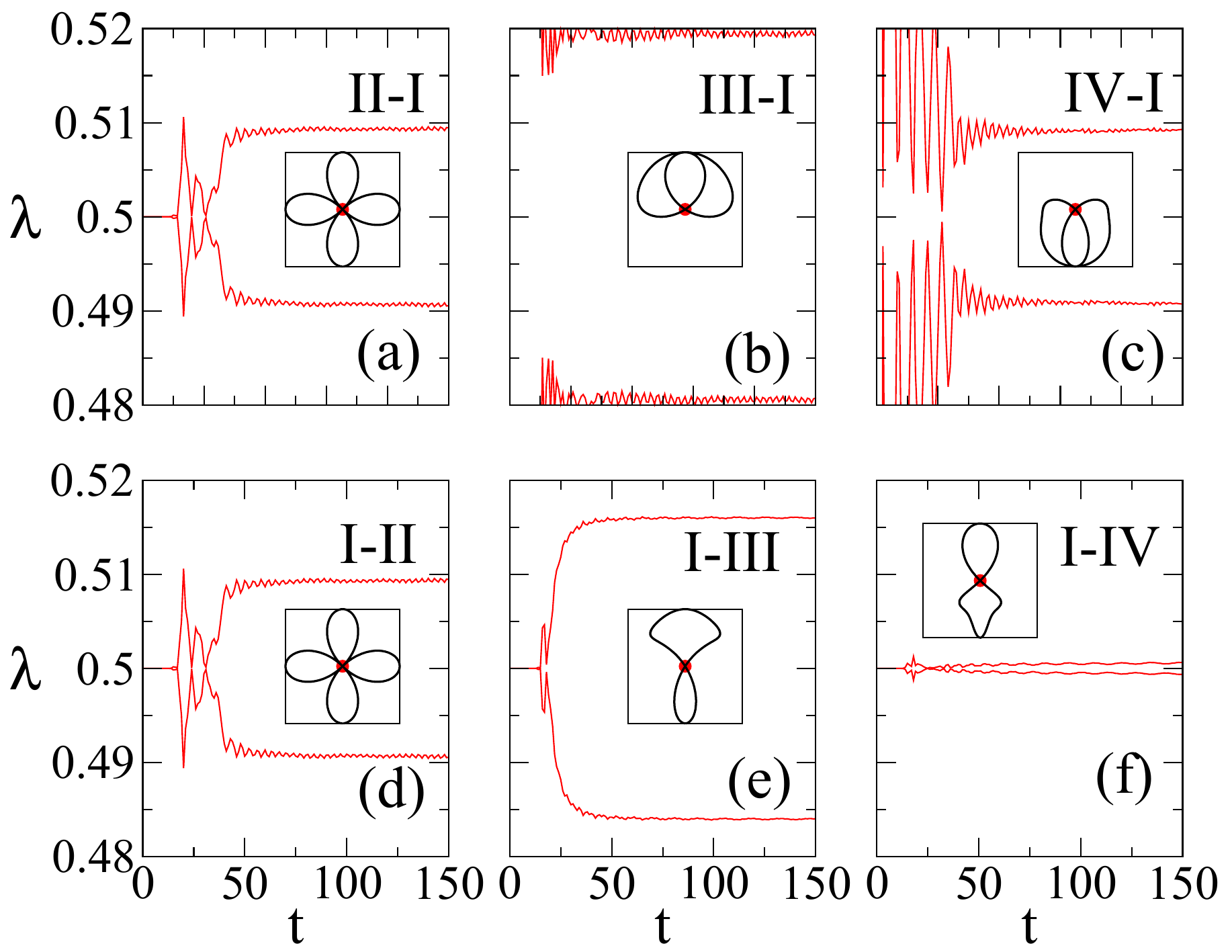}
\caption{(color online) Representative time evolutions of the OPES 
close to $1/2$ for quenches between (a) II-I, (b) III-I, (c) IV-I, 
(d) I-II, (e) I-III, and (f) I-IV. The initial or final parameters 
($\tilde{\mu}/2$, $\tilde{\Delta}$) are (0.5,2) for phase I, 
(0.5,-2) for phase II, (2, 2) for phase III, and (-2.2) for phase IV. 
The size of the subsystem A is $L=100$. 
Insets: The corresponding curves of ${\bs R}_{\mbox{\text eff}}(k)$ in 
the $R_y$-$R_z$ plane. The absence of the tMES in the 
long-time limit is reflected by the passing of 
${\bs R}_{\mbox{\text eff}}(k)$ at the origin.}
\label{fig2}
\end{figure}

Figs.~\ref{fig2}(a)-(c) show the time evolution of the OPES near $1/2$
by suddenly quenching the systems from phases II, III, and IV,
respectively, to phase I. We find that the Majorana zero-modes fail to
appear after a sufficiently long time, regardless of the topological
properties of the original state. In other words, the quench of the
topological systems with the Majorana edge modes cannot be
thermalized. For comparison, Figs.~\ref{fig2}(d)-(f) show the time
evolution of the OPES for the sudden quench from phase I to phases II,
III, and IV, respectively. The degenerate eigenvalues $\lambda_m =
1/2$ persist for some time before they split and relax to separate
values, depending on the final Hamiltonian. In other words, the
Majorana zero-modes before the quench are destroyed eventually.

What determines the finite survival time of the degenerate $\lambda_m
= 1/2$ in Figs.~\ref{fig2}(d)-(f)? We can imagine that upon quench
quasiparticles are generate in the bulk and propagate at a maximum
velocity $v_{\rm max} = [\partial \epsilon(k) / \partial k ]_{\rm
  max}$. Therefore, at $T^{\star} = L/(2 v_{\rm max})$ the Majorana
zero-modes at the boundaries of the entanglement cut can exchange
information and hence the degenerate levels in the OPES start to
split. Chaotic oscillations then emerge in the entanglement spectrum
due to the complex processes of quasiparticle interference and
decoherence.

To confirm the topology of the steady states of the quench process
in the infinite-time limit, we calculate the time-dependent
pseudomagnetic field ${\bs R}(k,t)$ from the time-dependent CFM
$G(k,t)$ in the Fourier space through the relation $G(k,t) = \left[ 1
  - {\bs R}(k,t) \cdot {\bs \sigma} \right] / 2$. We find that ${\bs
  R}(k,t) = \cos{(4 R' t)} \hat{\bs R}(k) + \sin{(4 R' t)} \hat{\bs
  R}(k) \times \hat{\bs R}'(k) + \left[ 1-\cos{(4 R' t)} \right] \left
[\hat{\bs R}(k) \cdot \hat{\bs R}'(k) \right ] \hat{\bs R}'(k)$, where
$\hat{\bs R}(k) \equiv {\bs R}(k) /R$ and $\hat{\bs R}'(k) 
\equiv {\bs R}'(k) /R'$.
In the infinite-time limit the sinusoidal time dependence dephases
away and $G(k,t = \infty)$ depends only on the effective
pseudomagnetic field $ {\bs R}_{\mbox{\text eff}}(k) \equiv \left
[\hat{\bs R}(k) \cdot \hat{\bs R}'(k) \right ] \hat{\bs R}'(k)$.
Therefore, the topology of the steady state is
determined by both the initial and final Hamiltonians: the quench
dynamics has a memory of the initial Hamiltonian, albeit entangled
with the final Hamiltonian. The insets in Fig.~\ref{fig2} dipict ${\bs
  R}_{\mbox{\text eff}}(k)$ in the above cases. We confirm that all
the traces pass through the origin, indicating that the Majorana
zero-modes are not stable in the infinite-time limit.

\begin{figure}
\center
\includegraphics[width=8cm]{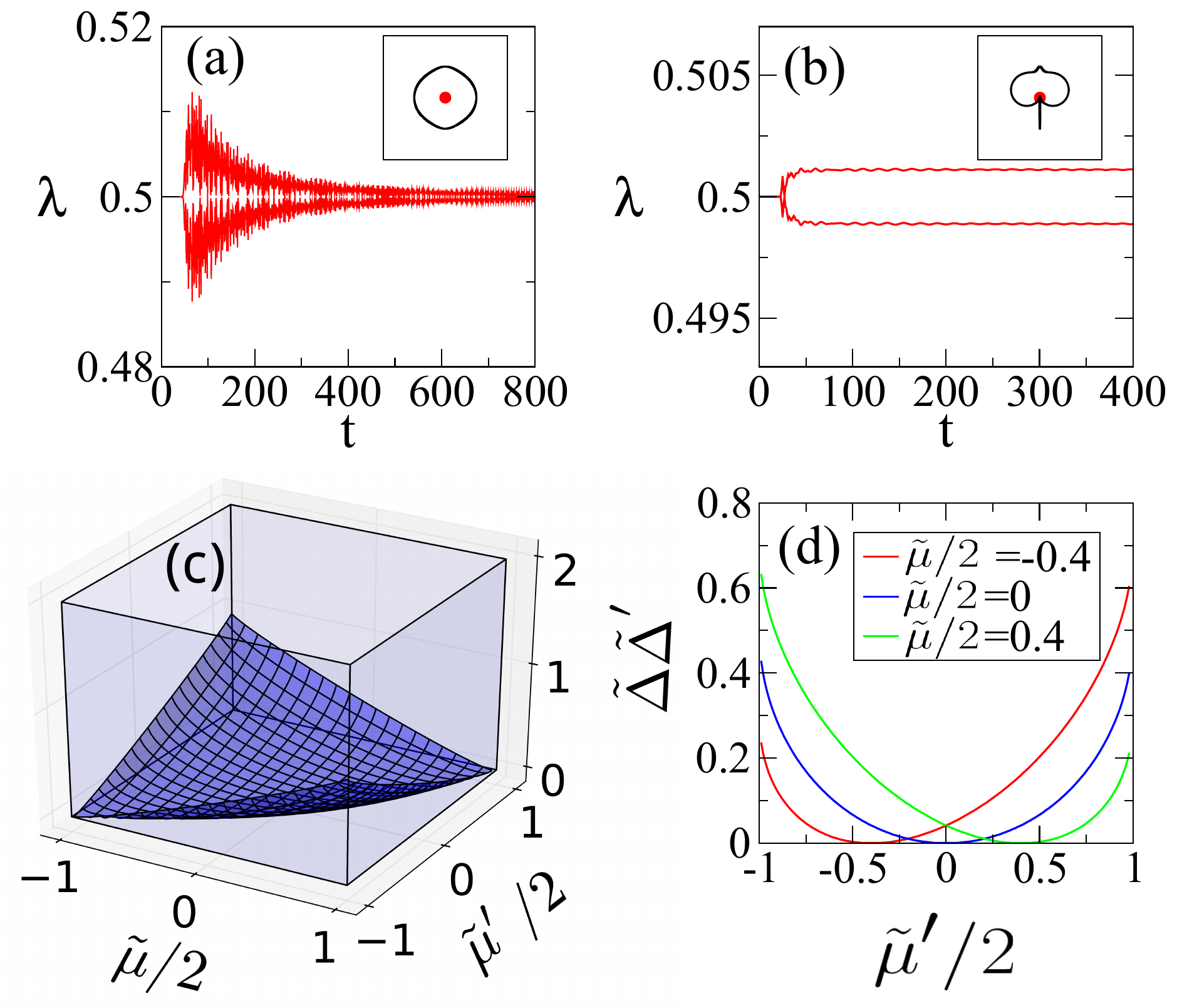}
\caption{(color online) The time evolution of the OPES near $1/2$
for two sudden quenches [marked by (a) solid arrow and 
(b) dashed arrow in Fig.~\ref{fig1}] within phase I. 
The tMES are recovered eventually in (a), but not in (b). 
Insets in (a) and (b):  Traces of ${\bs R}_{\mbox{\text  eff}}(k)$ in the 
$R_y$-$R_z$ plane. 
(c) The critical surface in the space of 
$\tilde{\mu}/2$, $\tilde{\mu'}/2$, and 
$\tilde{\Delta}\tilde{\Delta'}$ 
above which ${\bs R}_{\mbox{\text  eff}}(k) = 0$ has no solution. 
(d) Cross sections of the 
critical surface at $\tilde{\mu}/2 = -0.4$, 0, and 0.4.  
The size of subsystem A is $L=100$. }  
\label{fig3}
\end{figure}

{\it Quench within a topological phase--} Will Majorana zero-modes
persist if we quench between two Hamiltonians in the same topological
phase? Given the fact that the edge states cannot thermalize, the
naive answer of yes needs to be examined. In Figs.~\ref{fig3}(a) and
(b) we show the OPES evolution near 1/2 for two quantum quenches both
within phase I, as indicated in the phase diagram in
Fig.~\ref{fig1}. Surprisingly, we find that the Majorana zero-modes
reappear in the steady state in the infinite-time limit in
Fig.~\ref{fig3}(a), while disappear in Fig.~\ref{fig3}(b). This
contrasts to the persistence of the edge modes in a dimerized
chain~\cite{MJCM13}. We also plot the corresponding ${\bs
  R}_{\mbox{\text eff}}(k)$ in the insets of Fig.~\ref{fig3}(a) and
(b). ${\bs R}_{\mbox{\text eff}}(k)$ encircles the origin in the
former case, which confirms the persistent memory of the Majorana
modes after the quench. In sharp contrast, ${\bs R}_{\mbox{\text
    eff}}(k)$ passes through the origin in the latter case, which is
consistent with the loss of the memory of the Majorana modes. 
Note that the quench processes within phase II are similar. 

We conclude from the comparison that even for a sudden quench within
the same topological phase it is possible that $\hat{\bs R}(k) \cdot
\hat{\bs R}'(k) = 0$, such that the pseudomagnetic field ${\bs
  R}_{\mbox{\text eff}}(k)$ vanishes.  Explicitly, this means 
\be
\label{eq:criterion}
\left (\cos{k} + \frac{\langle \tilde{\mu} \rangle}{2}\right)^2  
+ \tilde{\Delta} \tilde{\Delta}' \sin^2{k} = 
\left (\frac{\delta\tilde{\mu}}{4}\right)^2 , 
\ee 
where $\langle \tilde{\mu} \rangle = (\tilde{\mu} + \tilde{\mu}')/2$ 
and $\delta\tilde{\mu} = \tilde{\mu} - \tilde{\mu}'$. 
Fig.~\ref{fig3}(c) shows the critical surface below
which the equality can be satisfied for some $k$; 
above the surface ${\bs R}_{\mbox{\text
    eff}}(k)$ encircles the origin (to ensure that both the initial
and final Hamiltonians are in the topological phase, we also need $\vert
\tilde{\mu} \vert, \vert \tilde{\mu}' \vert \le 2$), hence the
Majorana zero-modes are memorized in the long-time limit. 
This can be understood as follows: 
if the superconducting gaps of the initial and final Hamiltonians 
have no energy overlap, the Majorana modes are suppressed 
due to the mismatch of the corresponding single-particle states. 
In Fig.~\ref{fig3}(d) we plot the cross sections of the 
critical surface at $\tilde{\mu}/2 = -0.4$, 0, and
0.4; evidently, the minima of the product
$\tilde{\Delta}\tilde{\Delta}'$ are zero at $\tilde{\mu} =
\tilde{\mu}'$ when the gaps sit right on top of each other. 
   
\begin{figure}
\center
\includegraphics[width=8cm]{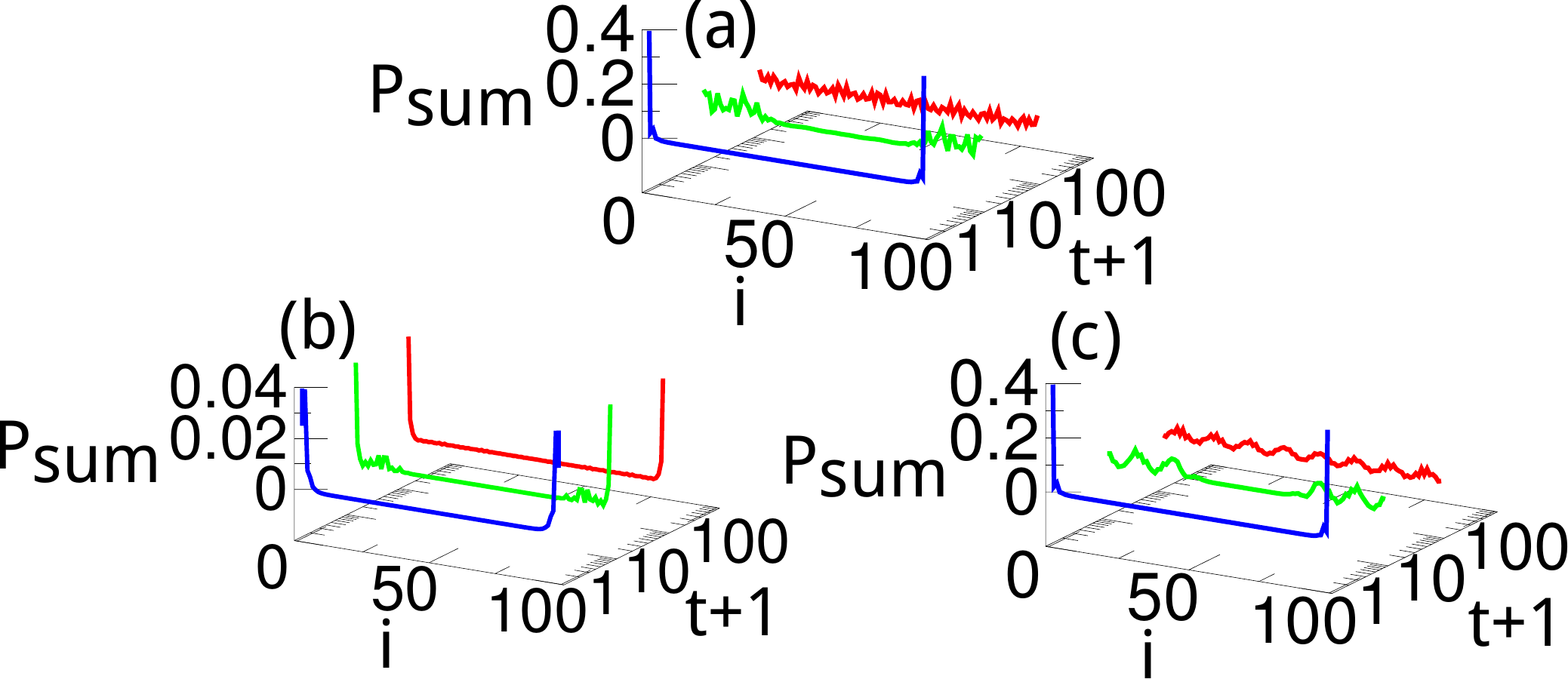}
\caption{(color online) The probability sum 
$P_{sum} = \sum_{j = 1,2} \vert \psi_j \vert^2$
of the two eigenstates of the entanglement Hamitonian 
whose eigenvalues are closest to $1/2$ 
at $t = 0$, 9, and 99
after a quench (a) from I (0.5,2) to II (0.5,-2), 
(b) from I (0.5,2) to I (0.5,1) as in Fig.~\ref{fig3}(a), 
and (c) from I (0.5,2) to I (-0.5,0.1) as in Fig.~\ref{fig3}(b).} 
\label{fig4}
\end{figure}

To support our findings, we analyze the eigenstates of the
entanglement Hamiltonian in the time evolution after the quench.  In
Fig.~\ref{fig4} we plot the probability sum $P_{sum} = \sum_{j = 1,2}
\vert \psi_j \vert^2$ of the two states $\psi_{1,2}$ whose eigenvalues
are closest to $1/2$ at $t = 0$, 9, and 99. As expected, $P_{sum}$
exhibits sharp peaks due to the presence of the Majorana edge
modes. The only case that such peaks survive [Fig.~\ref{fig4}(b)] is
the quench process within phase I as discussed in
Fig.~\ref{fig3}(a). For comparison, such peaks dissolve into the bulk
when we quenches the system to a different topological phase
[Fig.~\ref{fig4}(a)] or within phase I but with mismatching
superconducting gaps [Fig.~\ref{fig4}(c)].

In summary, we study the quench dynamics of a 1D p-wave superconductor
using the OPES. We find that the system reach a final steady state 
whose topology can be determined by an effective pseudomagnetic
field ${\bs R}_{\mbox{\text eff}}(k)$. As expected, sudden quenches from 
a topological phase to a trivial phase destroy the Majorana edge modes. 
However, the memory of the Majorana modes will also be lost if we quench
the system to a different topological phase, or if the superconducting 
gaps before and after the quench do not match. When both topological and 
energetic criteria are satisfied, the Majorana zero-modes will return 
after sufficiently long time.

MCC, PCC, and CYM acknowledge the NSC support under the contract Nos.
102-2112-M-005-001-MY3, 101-2112-M-007-010, and 100-2112-M-007-011-MY3,
respectively.  
XW acknowledges the support by the 973 Program under Project
No. 2012CB927404 and the NSFC Project No. 11174246.

\end{document}